\documentclass[preprint,preprintnumbers,amsmath,amssymb]{revtex4}

\usepackage{graphicx}
\usepackage{dcolumn}
\usepackage{bm}

\begin{document}

\title{Investigation of the $^{19}$Na nucleus via resonance elastic scattering}

\author{B.B.~Skorodumov}
\email{bskorodo@nd.edu}
\affiliation{Institute of Structure and
Nuclear Astrophysics, University of Notre Dame, Notre Dame, IN
46556}

\author{G.V.~Rogachev}
\affiliation{Institute of Structure and Nuclear Astrophysics,
University of Notre Dame, Notre Dame, IN 46556}
\affiliation{Physics Department, Florida State University,
Tallahassee, FL 32306}

\author{P.~Boutachkov}
\affiliation{Institute of Structure and Nuclear Astrophysics,
University of Notre Dame, Notre Dame, IN 46556}

\author{A.~Aprahamian}
\affiliation{Institute of Structure and Nuclear Astrophysics,
University of Notre Dame, Notre Dame, IN 46556}

\author{J.J.~Kolata}
\affiliation{Institute of Structure and Nuclear Astrophysics,
University of Notre Dame, Notre Dame, IN 46556}

\author{L.O.~Lamm}
\affiliation{Institute of Structure and Nuclear Astrophysics,
University of Notre Dame, Notre Dame, IN 46556}

\author{M.~Quinn}
\affiliation{Institute of Structure and Nuclear Astrophysics,
University of Notre Dame, Notre Dame, IN 46556}

\author{A.~Woehr}
\affiliation{Institute of Structure and Nuclear Astrophysics,
University of Notre Dame, Notre Dame, IN 46556}

\date{\today}

\begin{abstract}

The structure of the unbound proton-rich isotope  $^{19}$Na was
studied in resonance elastic scattering of a radioactive $^{18}$Ne
beam on a proton target using the thick-target inverse-kinematics
method. The experiment covered from 0.5 meV to 2.7 meV in c.m.s.
Only one state of $^{19}$Na (the second excited state) was
observed. A combined \textit{R}-matrix and potential model
analysis was performed. The spin and parity assignment of this
second excited state was confirmed to be 1/2$^{+}$. We show that
the position of the 1/2$^{+}$  state significantly affects the
reaction rate through that state but the total reaction rate
remains unchanged since the $^{18}$Ne(2p,$\gamma$) proceeds mostly
via the ground and first excited states in $^{19}$Na at stellar
temperatures.

\end{abstract}

\maketitle

\section{Introduction}

The structure of light exotic isotopes is currently one of the
major topics in nuclear physics. Properties of light
neutron-deficient isotopes are not as well known as the properties
of their neutron-rich mirrors. The Coulomb interaction leads to
proton unbound states in neutron-deficient nuclei even though the
corresponding states are bound in the mirror nuclei. Decay
properties of these nuclei yield additional information about
their structure as well as the structure of states in the mirror
nuclei. One such example is $^{19}$Na, a proton drip-line isotope
which is unbound with respect to proton emission by only 321 keV.
It was first observed by Cerny et al. \cite{Cer69} and until very
recently only its mass and the excitation energy of its first
excited state were experimentally known. The lack of experimental
information on the level structure of $^{19}$Na is due to
technical difficulties related to the population of the excited
states of this rather exotic nucleus with three fewer neutrons
than the stable sodium isotope $^{22}$Na. The second excited state
in $^{19}$Na was recently observed in a $^{18}$Ne+p elastic
scattering experiment \cite{Ang03} in which a narrow excitation
energy range (0.4 - 1.1 MeV) was covered. The original goal of
this work was to extend our knowledge of the structure of
$^{19}$Na to higher excitation energies (up to 2.7 meV) using the
resonance elastic scattering of a radioactive beam of $^{18}$Ne on
protons (the upper limit in excitation energy is implied by the
maximum possible energy of $^{18}$Ne beam that can be produced at
ISNAP \cite{Lee99}). During the review stage of this manuscript
new study of the $^{19}$Na nucleus was published from the study at
SPIRAL (GANIL) by Oliveira et al. \cite{Oli05}. In this work, the
spectrum of $^{19}$Na was measured up to excitation energy of 5.7
MeV at one angle (180$^\circ$) using the Thick Target Inverse
Kinematics (TTIK) technique \cite{Art90}. Two additional
resonances at higher excitation energies were found in
\cite{Oli05}. In addition, two broad peaks  at $E_{cm}\approx$ 2.4
and 3.1 MeV were observed. It was pointed out by the authors in
\cite{Oli05} that these peaks can originate from process other
than elastic scattering. The first peak is in the energy range of
our experiment and should be populated if the origin of this peak
is in fact related to elastic scattering. Therefore, it is of
interest to compare our data, obtained at a lower beam energy and
different angles with the data from \cite{Oli05}. The structure of
low lying resonances in $^{19}$Na also has some astrophysical
interest. It was shown in \cite{Gor95} that at certain stellar
temperature and density conditions two-proton capture reaction on
$^{18}$Ne can play an important role in bridging the waiting point
nucleus $^{18}$Ne and provide continuous flow between the CNO and
the Fe-Ni mass region in the \textit{rp} -process. The
$^{18}$Ne(2p,$\gamma$) reaction rate was recalculated taking into
account new information on the excitation energy of the second
excited state.

In the following sections, we present the results of resonance
elastic scattering of $^{18}$Ne on protons, discuss the structure
of the $^{19}$Na nucleus, and finally discuss the influence of a
$1/2^+$ state on two-proton capture reaction rate on the $^{18}$Ne
nucleus.

\section{\label{sec:exp} Experiment}

\begin{figure}[]
\includegraphics[width=12.0cm]{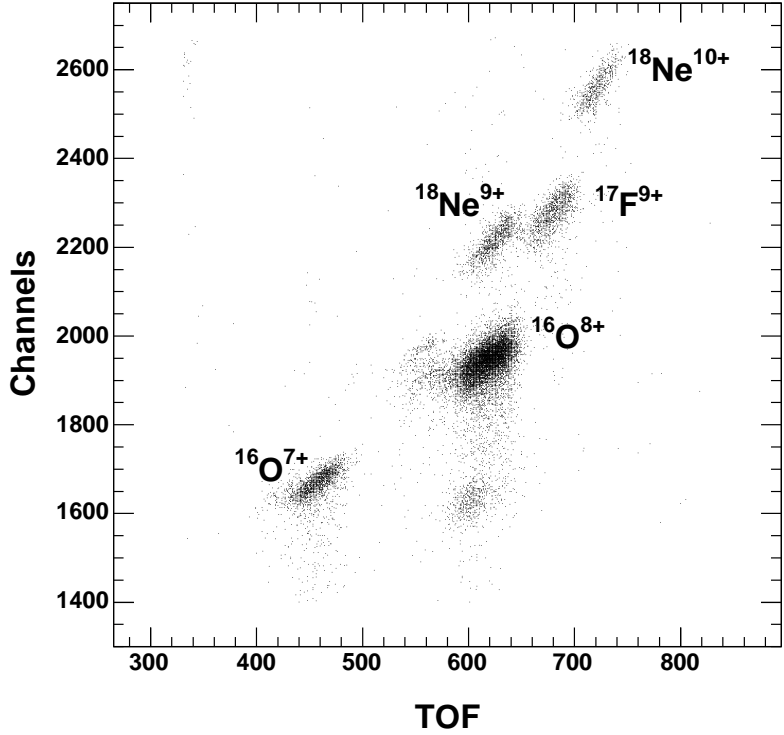}
\caption{\label{fig:timing}Composition of the radioactive beam
measured by a Si detector placed at the secondary target position.
The energy of the ions is shown versus the time of flight (ToF).
The ``Start'' signal for the ToF measurements was produced by the
Si detector, and the ``Stop'' signal was derived from the
radiofrequency (RF) signal of the buncher.}
\end{figure}

\begin{figure}[]
\includegraphics[width=12.0cm]{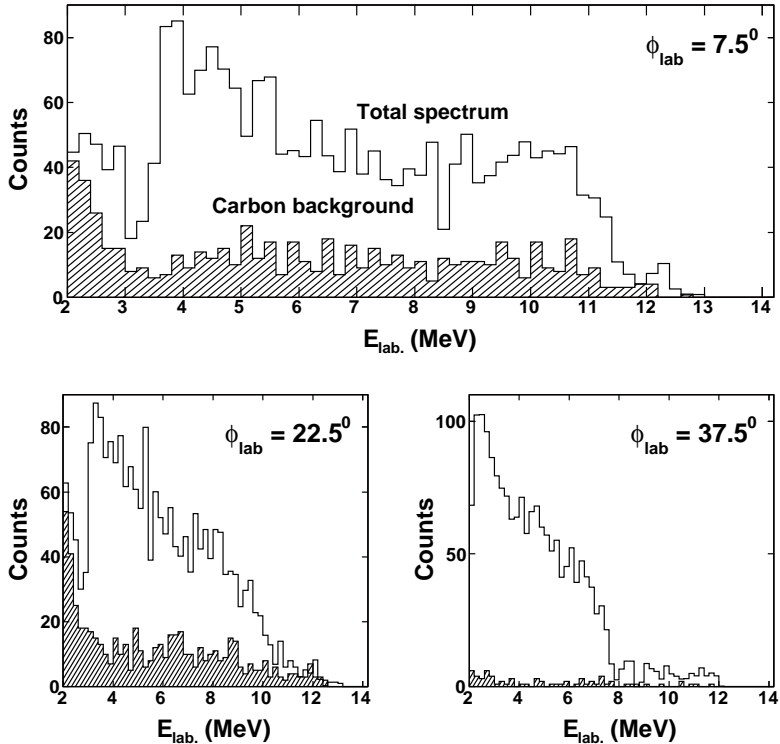}
\caption{\label{fig:lab}The spectra of protons from the
polyethylene target observed in Si detectors placed at
7.5$^\circ$, 22.5$^\circ$ and 37.5$^\circ$ in the laboratory
frame. The shaded areas represent the proton spectra measured with
a natural carbon target. The spectra were gated only on the ToF
signal of the $^{18}$Ne$^{10+}$ beam (see Fig. \ref{fig:timing}).}
\end{figure}

The experiment was carried out at the TwinSol radioactive nuclear
beam facility \cite{Lee99} of ISNAP (Institute of Structure and
Nuclear Astrophysics) at the University of Notre Dame. A beam of
$^{18}$Ne was produced via the $^{3}$He($^{16}$O,$^{18}$Ne)n
reaction. The experimental setup is shown in Fig. \ref{fig:setup}.
A primary beam of $^{16}$O with an average intensity of 230
electrical nA and an energy of 80 MeV was incident on a 2.5 cm
long gas cell containing $^{3}$He gas at a pressure of 1 atm. A
Faraday cup placed after the gas cell was used to stop the primary
beam and to measure its intensity. Two superconductive solenoids
acted as thick lenses to separate the $^{18}$Ne beam from other
reaction products. The magnetic rigidity of the first
superconductive solenoid was set to focus $^{18}$Ne$^{+9}$ ions
with an energy of $~$57 MeV onto a 10 $\mu$g/cm$^{2}$ C stripping
foil as shown in Fig. \ref{fig:setup}, while the second solenoid
focused fully-stripped $^{18}$Ne$^{+10}$ ions onto the secondary
target. The different rigidities of the two solenoids allowed for
a much better separation of $^{18}$Ne from other reaction
products. The $^{18}$Ne ions were focused onto a 1 cm (diameter)
spot at the secondary target position. Under these conditions, a
$^{18}$Ne beam was obtained with an intensity of about 10$^{4}$
pps, an energy of 56.3 MeV, and an energy spread of 3.2 MeV full
width at half maximum (FWHM). The desired $^{18}$Ne ions made up
only 2\% of the secondary beam; the main contamination was
$^{16}$O (see Fig. \ref{fig:timing}). Polyethylene (CH$_{2}$) foil
was used as the secondary target. Its thickness of (5.52
mg/cm$^{2}$) was chosen to stop the $^{18}$Ne ions completely. The
Thick-Target Inverse-Kinematics (TTIK) technique \cite{Art90} was
used to obtain an excitation function for resonance elastic
scattering. This method is based on the large differences of
specific energy losses for heavy ions ($^{18}$Ne) and protons.
Protons backscattered by $^{18}$Ne penetrate the target and lose
only a small fraction of their energy. The center-of-momentum
energy at the moment of the interaction can therefore be recovered
from the measured proton energy. Three Si detectors placed at
angles of 7.5$^\circ$, 22.5$^\circ$ and 37.5$^\circ$ with respect
to the beam axis (corresponding to  backward angles for elastic
scattering of $^{18}$Ne on protons) were used to detect recoil
protons. The primary $^{16}$O beam was pulsed with a repetition
rate of 5 MHz and a pulse width of approximately 2 ns. The 5.6 m
distance from the primary target to the array of silicon detectors
was adequate to obtain a clean separation of the protons
associated with the interaction of $^{18}$Ne ions in the target
from protons associated with elastic scattering of beam
contaminants and reactions of other beam isotopes in the target
(see Fig. \ref{fig:timing}). Spectra of protons (in the laboratory
frame) measured in the three detectors are shown in Fig.
\ref{fig:lab}. Measurements with a thick natural carbon target
were performed to subtract the background associated with carbon
in the polyethelyne target. The thickness of the carbon target was
matched to that of the polyethylene target in terms of proton
energy losses. The shaded areas in Fig. \ref{fig:lab} show the
contributions from reactions of $^{18}$Ne on the carbon target.

\begin{figure}[]
\includegraphics[width=12.0cm]{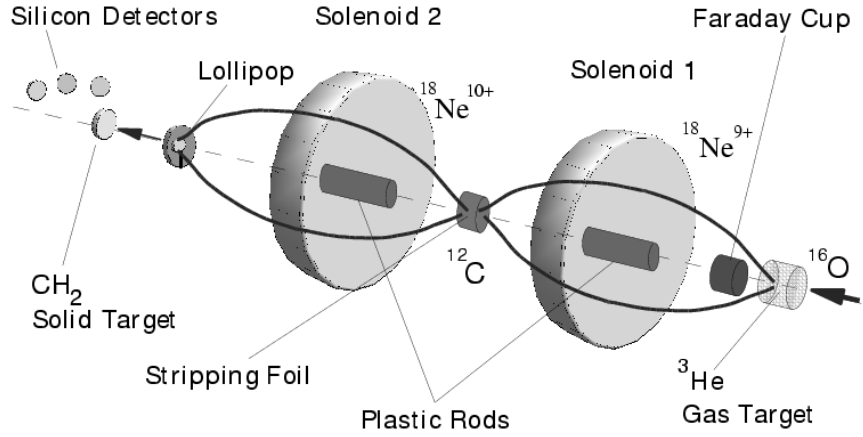}
\caption{\label{fig:setup}The experimental setup with beam
trajectories through TwinSol.  The "lollipop" reduces
contamination of the beam by intercepting ions that focus at a
different location relative to the $^{18}$Ne beam.}
\label{fig:setup}
\end{figure}

\begin{figure}[]
\includegraphics[width=12.0cm]{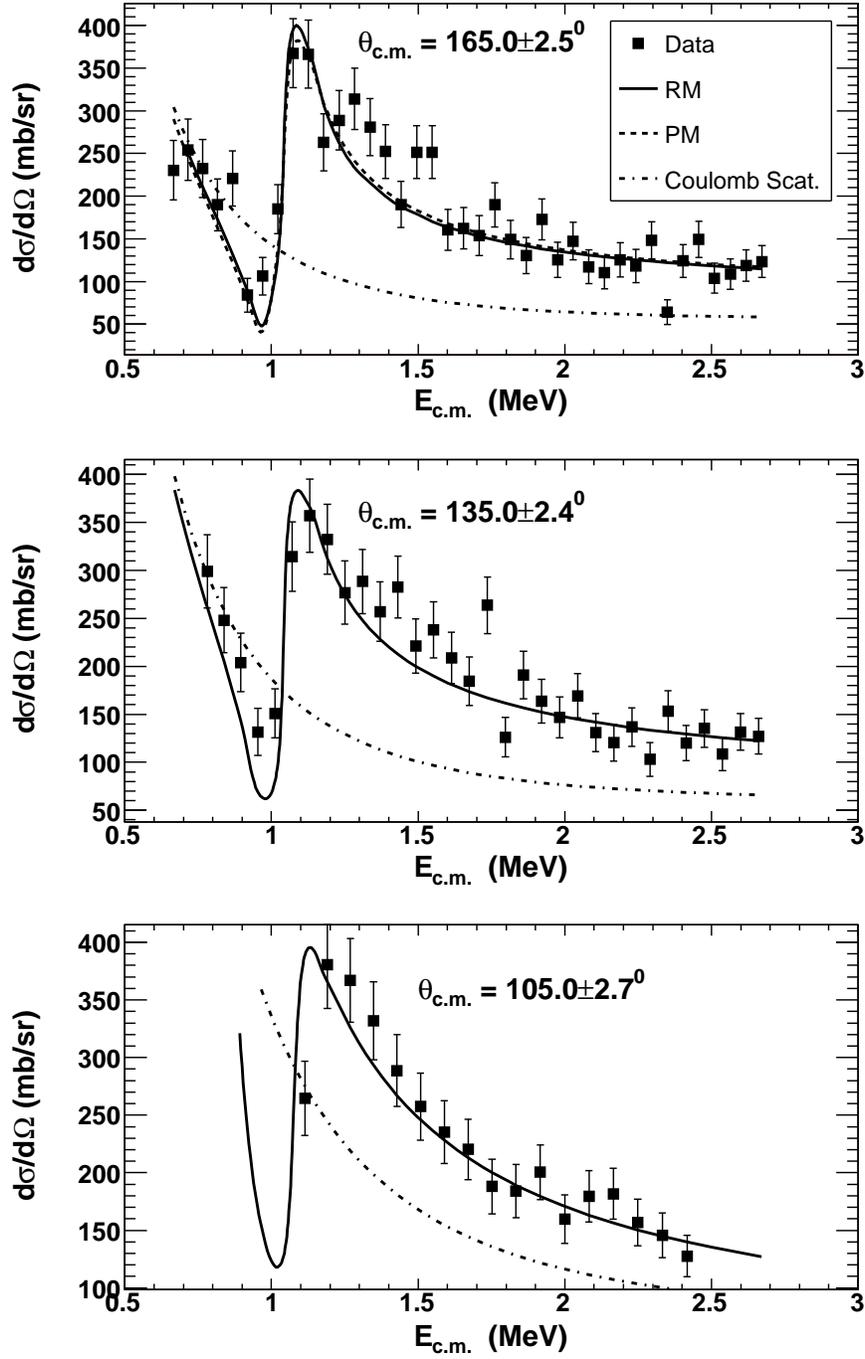}
\caption{\label{fig:exfun} The excitation functions for
$^{18}$Ne+p elastic scattering. The solid curves correspond to
\textit{R}-Matrix calculations, the dashed curve(only top figure)
to potential-model calculations, and the dashed-dotted curve to
Coulomb scattering.}
\end{figure}

Proton spectra taken at three different angles, converted to the
c.m. system, are shown in Fig. \ref{fig:exfun}. In the conversion
of raw spectra to the absolute cross section we used the measured
ratio of $^{18}$Ne to the intensity of the primary beam $^{16}$O.
The kinematic relations and energy losses of protons in the target
were taken into account in the transformation to the c.m.
excitation functions. Monte Carlo simulations of the experimental
energy resolution were performed for several excitation energies
and angles. The typical experimental resolution at 165$^{\circ}$
was found to be about $30$ keV at c.m. energies above 1 MeV. The
precision of the excitation energies is about $20$ keV, defined
mainly by the uncertainties in the specific energy losses of
$^{18}$Ne ions in the CH$_{2}$ target. The absolute cross section
values are obtained with uncertainties of 15\%. Only one resonance
was observed in the $^{18}$Ne+p elastic-scattering excitation
function in the measured energy range up to 2.7 MeV. This
resonance was previously reported in \cite{Ang03} and
\cite{Oli05}. We confirm the 1/2$^{+}$ spin-parity assignment for
the resonance. The interference pattern of the $l=0$ resonance
phase-shift with Coulomb phase-shift and the angular distribution
(three different angles were measured) provides for unique
spin-parity assignment $1/2^+$ for the observed resonance. The
excitation energy and width of this state were obtained in an
\textit{R}-Matrix fit: $E_{ex}$= 0.74 $\pm$ 0.03 MeV and $\Gamma$
= 130 $\pm$ 50 keV. Statistical uncertainties dominate the
reported errors.  The parameters of the resonance are in agreement
with the results obtained for this state in \cite{Ang03}.
Continuum Shell Model (CSM) calculations \cite{Vol04}, performed
for a chain of oxygen isotopes, predict that the 1/2$^{+}$ second
excited state in $^{19}$O (the mirror of the observed state in
$^{19}$Na) has an almost pure ($\approx$ 70\%) single particle
nature. We need a proper determination of the single particle
spectroscopic factor for this resonance to test the predictions of
the CSM. Often, the Wigner limit $ \gamma_{W}^{2} =
\frac{3\hbar^2}{2\mu r^2}$ is used to estimate the single particle
width of the resonance and to extract a spectroscopic factor.
However, this approach gives only an order-of-magnitude estimate.
When applied to the observed 1/2$^{+}$ state, a misleading result
is obtained. In particular, the single particle spectroscopic
strength extracted for the state is only $~$ 25\% \cite{Ang03}. In
this work, we use the potential-model approach instead. In this
case, the properties of the $^{18}$Ne+p system are described by a
spherical Woods--Saxon potential. If the parameters of the
potential are known, then one can predict the properties of the
single particle states. The width of the states calculated with
the proper potential can then be compared with the experimental
values and the spectroscopic factor can be extracted as the ratio
of experimental width to the calculated value (as long as only one
decay channel is open, as is the case in the energy region of
interest). The same approach was adopted in \cite{Oli05} but no
details on the potential parameters are given. The initial
parameters of the Woods--Saxon potential were found by fitting the
excitation energies of the lowest 5/2$^{+}$, 1/2$^{+}$, and
3/2$^{+}$ levels in $^{17}$O and $^{17}$F simultaneously
\cite{Gol04}. It is known that these states are pure single
particle in nature. (The Spectroscopic Factors (SF) are 0.9, 0.9
and 1.25, respectively, \cite{Til93}). A single set of
Woods--Saxon potential parameters generates correct energies for
the ground and the first two excited states in both $^{17}$O and
$^{17}$F to within 100 keV. The optimum parameters are shown in
the table \ref{tab:levels}. The width of the 3/2$^{+}$ state (the
only unbound state among the three in $^{17}$O/$^{17}$F)
calculated with the potential given in the table \ref{tab:levels}
is 90 keV for $^{17}$O and 1.14 MeV for $^{17}$F, \cite{Til93}.
The experimental values are 95 keV and 1.5 MeV, respectively. The
ratio of the experimental to the calculated width gives the
spectroscopic factor, which is greater than one for both $^{17}$O
and $^{17}$F. To avoid this, we could increase the diffuseness of
the potential ( as it was suggested in \cite{Gol04}) to get SF=1.0
for 3/2$^+$ instead of 1.25. But since we want to keep all
potential parameters the same for $^{17}$O/$^{17}$F and
$^{19}$O/$^{19}$Na (except V$_{0}$), we decided to use standard
values. A potential with the same Woods--Saxon parameters (except
for V$_{0}$ parameter) was used to calculate the widths and
excitation energies of the single particle states in $^{19}$O. The
potential was adjusted to fit the position of the first 1/2$^{+}$
state in $^{19}$O which has pure single particle structure
\cite{Til95} (see Table \ref{tab:levels}). The excitation energy
of the other pure single particle state (3/2$^{+}$at 5.45 MeV) was
reproduced by this potential with an accuracy of 40 keV. The
potential model predicts the width of this latter state to be 330
keV, which is consistent with the experimental value of 320 $\pm$
25 keV \cite{Til95}. The ground state of $^{19}$O appears to be
underbound by 500 keV, but it is known from shell model
calculations \cite{Vol04} and also from experiment \cite{Til95}
that the $^{19}$O ground state has significant admixtures from
configurations other than simple ($d_{5/2}$)$^{3}$ for the
neutrons in excess of \textit{N} = 8. Thus, our potential model is
not adequate for a description of the ground state of $^{19}$O.
The potential obtained in the procedure described above can now be
used to calculate the single-particle width of the 1/2$^{+}$ state
in $^{19}$Na. It was found that the predicted width of this state
is 134 keV, in comparison with the experimental value of 130 $\pm$
50 keV gives a clear indication that it is a pure single particle
state (SF = 0.95 $\pm$ 0.3). The spectroscopic factor defined in
this way is in very good agreement with the result of CSM which
gives 70\% SF and a width for the 1/2$^+$ resonance in $^{19}$Na
of about 94 keV (obtained as follows: the single particle width
from potential model was multiplied by SF given by CSM). The
excitation function for $^{18}$Ne+p elastic scattering can be
calculated using the potential with the parameters given in the
table \ref{tab:levels}. The dashed curve in Fig. \ref{fig:exfun}
shows the result of such a calculation and it is almost identical
to the \textit{R}-matrix fit, shown with the solid curve.

\begin{table}
\caption{\label{tab:levels}Parameter of the Woods--Saxon
potential}
\begin{tabular}{|m{2.6cm}|m{2.4cm}|m{2.5cm}|}
\hline Parameter& $^{17}$O /$^{17}$F &
$^{19}$O/$^{19}$Na\tabularnewline \hline $V_{0}$& -55.77&
-50.12\tabularnewline $V_{0sl}$& 6.40& 6.40\tabularnewline
$r_{0}$& 1.21 & 1.21\tabularnewline $r_{0sl}$& 1.21&
1.21\tabularnewline $r_{Coul}$& 1.21& 1.21\tabularnewline $a$&
0.65& 0.65\tabularnewline $a_{sl}$& 0.65& 0.65\tabularnewline
\hline
\end{tabular}
\end{table}

\begin{figure}[]
\includegraphics[width=12.0cm]{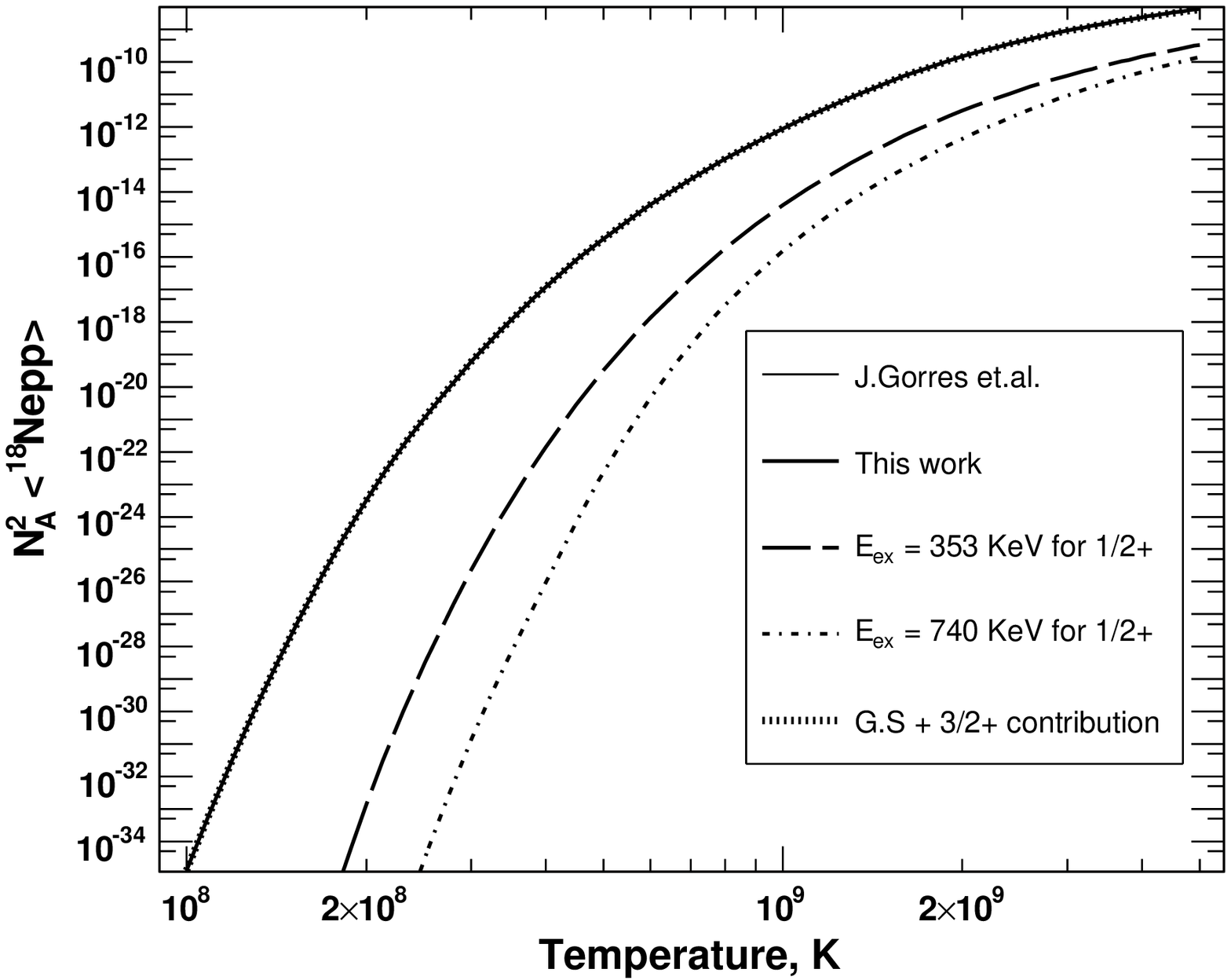}

\caption{\label{fig:astro} The reaction rate for two-proton
capture $^{18}$Ne(2p,$\gamma$). The calculations were performed in
noneresonant and resonant analytical approach used in
\cite{Gor95}. The thin, thick and the slashed lines overlaps among
each other which show reaction rate for two proton capture on
$^{18}$Ne which includes the contribution from ground state
(5/2$^+$ in $^{19}$Na ) and the first two excited states (3/2$^+$,
1/2$^+$ in $^{19}$Na) except slashed line in which excluded
contribution from 1/2$^+$ state in $^{19}$Na. Dashed and
dot-dashed lines present contribution from the $1/2+$ level to the
two-proton capture with old and new position of $1/2+$ levels,
respectively.}
\end{figure}

The same procedure can be used to obtain the SF of the 1/2${^+}$
state from the experimental data of \cite{Ang03} by Angulo et al.,
where the low-energy part ($E_{cm}$ < 1.4 MeV) of the $^{18}$Ne+p
elastic scattering excitation function was measured. The width of
the 1/2$^{+}$ state was found to be $\Gamma$ = 100 $\pm$ 8 keV
\cite{Ang03}. (The error bars include the uncertainty in the
arbitrary choice of the channel-radius parameter in the
\textit{R}-matrix fit). The fit to the data from \cite{Ang03} and
our potential model gives a spectroscopic factor of 0.75 $\pm$
0.06. The error reflects the statistical uncertainty of the data
from \cite{Ang03}. The state observed at 2.37 MeV in $^{19}$O does
not manifest itself in the measured excitation function.
Nevertheless, conclusions can be drawn from the fact that we did
not see any indications of the presence of this state in our data.
Specifically, this state is either very narrow and/or only weakly
populated in the elastic channel. The CSM calculations indicate
that the third excited state of $^{19}$O/$^{19}$Na, is built
predominantly on the first and the second excited states of
$^{18}$O/$^{18}$Ne, and has a spin-parity of 9/2$^{+}$. The width
of this state in $^{19}$Na, according to the CSM prediction
\cite{Vol04} is 1.1 eV. However, this model does not take into
account possible admixtures from $g_{9/2}$ shell (only sd shell is
included). The single-particle width of this state calculated with
the potential given in Table \ref{tab:levels} is 0.5 keV. So even
if we assume the $10\%$ admixture of the $g_{9/2}$ shell to the
wave function of this state the width would become $\sim$ 5 eV.
Much too narrow to be measured in the present work. Hence, the
absence of the state around 2.4 MeV (excitation energy) in the
observed $^{18}$Ne+p spectrum lends indirect support to the
prediction of the CSM regarding the nature of this state.

In \cite{Oli05}, two peaks with cross sections of $\approx$ 300
mb/sr were observed at cm energies of 2.4 and 3.1 MeV in the
excitation function of $^{18}$Ne+p elastic scattering. As shown in
Figure \ref{fig:exfun} no peak is present at 2.4 MeV in our
spectrum. It is a clear experimental proof that peaks observed in
\cite{Oli05} are not related to the $^{19}$Na states populated in
elastic scattering. Authors of \cite{Oli05} argue that these peaks
can be related to the sequential 2p-decay of the highly excited
states of $^{19}$Na. Absence of these peaks in our spectrum lends
indirect support to this hypothesis because due to lower initial
$^{18}$Ne beam energy in our experiment these highly excited
states would not be populated and thus the peaks at 2.4 MeV and
3.1 MeV cannot be observed.

Potential model with potential parameters given in Table
\ref{tab:levels} reproduces excitation energies (with 100 keV
accuracy) and widths of all states of single particle nature in
$^{17}$F, $^{17}$O and $^{19}$O. It also gives a very good
description of the $^{18}$Ne+p excitation function up to 2.7 meV
(c.m.) (dashed curve in Fig. \ref{fig:exfun}a). Using this model
one can make a prediction for the excitation energy and width of
the 3/2$^+$ (1d3/2)$^1$ single particle state in $^{19}$Na.
According to our calculations it should have an excitation energy
of 5.2 meV and width of $\sim$ 2 meV. This state is the analog of
the known 5.45 meV 3/2$^+$ single particle state in $^{19}$O
\cite{Til93}. It is also a feature of both CSM calculations
\cite{Vol04} and shell-model calculation using code Oxbash
\cite{Bro86} with the WBT  \cite{War92} interactions where it
appears at 5.53 meV. Thus this state should clearly be in
accessible range of \cite{Oli05} where the excitation function was
measured up to 5.7 meV. However, we saw no indication of this
state. It would be of great interest to locate this state in
$^{19}$Na in future measurements or understand the reason for its
disappearance.

It was shown in \cite{Gor95} that for high temperature and density
conditions, two-proton capture on $^{18}$Ne will compete with
$(\alpha,p)$ and $\beta^+$ decay rate. Such stellar condition
would allow a fast depletion of $^{18}$Ne and a fast leakage out
of the hot CNO cycle toward higher masses. In \cite{Gor95}, the
second excited state $1/2$+ in the  $^{19}$Na nucleus, was not
known. The position of the $1/2$+ state was estimated from
Thomas--Ehrman shift relative to the mirror $1/2+$ state in
$^{19}$O and the value $E_{ex}$ = 0.353 meV was used. It is more
than two times lower than the experimental value obtained by us
and as a result some changes in reaction rates may occurs. We
recalculated the $^{18}$Ne(2p,$\gamma$) reaction rate for the same
stellar conditions as in \cite{Gor95}. For the case discussed
here, the approach proposed in \cite{Fow67} can be used.
 Using the obtained energy
of the $1/2+$ state, the $^{18}$Ne(2p,$\gamma$) reaction rate was
reduced by a factor of 100 due to the new location of the 1/2$^+$
resonance (Fig. \ref{fig:astro}, the ordinate represents total two
proton reaction rate on $^{18}$Ne). However, the total reaction
rate is almost unaffected since at stellar temperatures, the
reaction rate is dominated mostly by g.s. and first excited state.

\section{\label{sec:con} Conclusion}

The structure of the proton-rich particle unstable nucleus
$^{19}$Na was investigated in the excitation energy region from
0.5 meV to 2.7 meV by means of resonance elastic scattering of a
radioactive beam of $^{18}$Ne on protons. Only one state, at an
excitation energy of 0.74 meV, was observed. Its spin and parity
were confirmed to be 1/2$^{+}$. A potential model approach was
used to obtain the single particle spectroscopic factor of this
resonance. We show that the state has a reasonably pure
single-particle structure (2s1/2)$^1$, in agreement with
shell-model calculations \cite{Vol04}. The absence of another
low-lying state (observed at an excitation energy of 2.37 meV in
the mirror nucleus $^{19}$O) in the $^{18}$Ne+p excitation
function is indirect confirmation of the shell-model prediction
that this is a 9/2$^{+}$ state built on an excited states of
$^{18}$Ne. As such, it would be very narrow and only weakly
populated in elastic scattering.

We did not observe a broad peak at 2.4 meV, found in \cite{Oli05},
which indicates that this peak is unrelated to elastic scattering
and may be the result of sequential 2p decay from the higher lying
excited states of $^{19}$Na (as argued in \cite{Oli05}). The
potential model with potential parameters obtained in this work
was used to predict the excitation energy and width of a 3/2$^+$
single particle state in $^{19}$Na. This state should be well in
the range of the experimental data of work \cite{Oli05}, but there
is no indication of this state. Additional measurements at higher
excitation energies (and different angles) are needed to locate
this resonance or understand the reason for the disappearance of
this state.

The $^{18}$Ne(2p,$\gamma$) astrophysical reaction rate was
recalculated using the new experimental information on the
excitation energy of the 1/2$^+$ state. The (2p,$\gamma$) reaction
rate due to this state is actually 100 times lower than it was
suggested in previous work \cite{Gor95}, however, due to the fact
that the $^{18}$Ne(2p,$\gamma$) reaction at stellar temperatures
is dominated by the ground and the first excited state this
finding did not make significant impact on the total reaction
rate.

\section{\label{sec:thanks} Acknowledgements}

This work was supported by National Science Foundation under Grant
numbers PHY01-40324, PHY02-44989 and PHY01-39950. The authors are
grateful to V.Goldberg, J.G$\ddot{o}$rres, A.Volya and M.Wiescher
for useful discussions and comments.

\end{document}